\documentclass[prd,floats,floatfix,showpacs,nofootinbib]{revtex4}
\usepackage{graphicx}
\usepackage{dcolumn}
\usepackage{bm}
\usepackage{graphics}
\usepackage{slashed}
\usepackage{amssymb}
\usepackage{natbib}
\usepackage{amsmath}
\usepackage{amsthm}
\usepackage{url}
\usepackage{color}
\usepackage{enumerate}
\usepackage{ulem}

\newcommand{\fracc}[2]{\frac{\textstyle{#1}}{\textstyle{#2}}}

\begin{document}

\title{Intrinsically symmetric cosmological model in the presence of dissipative fluids}

\author{Eduardo Bittencourt} \email{bittencourt@unifei.edu.br}
\author{Leandro Gomes}\email{lggomes@unifei.edu.br}
\author{Grasiele Santos}\email{gbsantos@unifei.edu.br}
\affiliation{Federal University of Itajub\'a, Itajub\'a, Minas Gerais 37500-903, Brazil}

\date{\today}

\begin{abstract}
Based upon the intrinsic symmetries approach to inhomogeneous cosmologies, we propose an exact solution to Einstein's field equations where the spatial sections are flat and the source is a non-perfect fluid such that the dissipative terms can be written in terms of spatial gradients of the energy density under a suitable choice of the coordinate system. It is shown through the calculation of the luminosity distance as a function of the redshift that the presence of such inhomogeneities may lead to an effective deceleration parameter compatible with either the standard $\Lambda$CDM model or LTB models depending on the choice of boundary conditions with no exotic matter. This fact is another evidence that different inhomogeneous models should be carefully investigated in order to verify which model may be compatible with observations and still be as close as possible to the standard model regarding the underlying assumptions, without resorting necessarily to exotic matter components.
\end{abstract}

\maketitle

\section{Introduction}
Several attempts to describe our actual clumped universe have been made by adopting fewer symmetries than the standard cosmological model, described by the Friedmann-Lema\^itre-Robertson-Walker (FLRW) class of geometries, which is invariant under a six-parameter group of isometries and whose surfaces of transitivity are three-dimensional spacelike hypersurfaces of constant curvature \cite{ellis_mac_marteens}. A number of inhomogeneous geometries have been constructed following this idea in order to give a more realistic view of cosmology \cite{steph03,kra}. Special attention is given to models dealing with averaging and backreaction issues \cite{wiltshire,kolb,buchert}, the exact Lema\^itre-Tolman-Bondi (LTB) models \cite{lemaitre,tolman,bondi} and Szekeres models \cite{krasinski}.

A particularly interesting technique to generate inhomogeneous cosmologies is the one developed by Szafron and Collins \cite{coll_79,coll1_79,coll2_79,coll_grg} in which exact solutions to the Einstein's field equations (EFE) are found through the so-called intrinsic symmetries approach, where only submanifolds of the whole space-time admit certain groups of isometries. When the fluid flow is irrotational, for instance, one could impose symmetries on the hypersurfaces orthogonal to the flow, seen as three-dimensional manifolds, and still the full space-time would possess no symmetries whatsoever.

Following this procedure, exact solutions were studied considering simple matter contents, as vacuum \cite{wolf86}, irrotational dust \cite{bona92,sop00} and perfect fluids \cite{arg85, wolf86a}. In Ref.\ \cite{wolf86}, flatness of three-dimensional hypersurfaces (that could be either timelike or spacelike) was imposed together with the condition of a traceless extrinsic curvature, for simplicity. In Ref.\ \cite{bona92}, irrotational dust metrics were considered by demanding that the hypersurfaces orthogonal to the fluid flow had constant curvature, and the authors found that the resulting solutions are either contained in the Szekeres dust solutions \cite{sze75} or are homogeneous of certain Bianchi types. Also, in Ref.\ \cite{sop00}, an irrotational dust solution was considered and a maximal group of isometries, that is, a six-parameter group of motions was imposed on the spacelike hypersurfaces orthogonal to the fluid velocity, and the conclusion was that all the irrotational dust solutions of EFE with flat spatial geometry were either Bianchi I or subfamilies of the Szekeres geometry.

In this work we propose an explicit inhomogeneous solution of the EFE by restricting the hypersurfaces orthogonal to the irrotational fluid flow to be flat, but allowing the matter content to have dissipative components. Assuming certain phenomenological thermodynamic equations, it is possible to completely determine the geometry and the matter distribution only by specifying the functional form of the energy density. The EFE will constrain the set of possibilities narrowing the phenomenological thermodynamic parameters and the spatial part of the energy density. Then, we show how the inhomogeneities in the model yield an effective deceleration parameter comparable to those found in the literature. In particular, compared to the FLRW model, this might lead to a behavior usually attributed to the dark energy component.

We begin by presenting in the next section the inhomogeneous solution in detail. Then, in Sec.\ \ref{III}, we move to the analysis of the luminosity distance and the calculation of the deceleration parameter in the approximation of small values of the redshift. We then proceed to put bounds, in Sec. \ref{IV}, on the parameters of our model based on the observed value of the deceleration parameter according to $\Lambda$CDM and predictions from LTB models. Concluding remarks are presented in Sec.\ \ref{V} and, for completeness, we compute the Newman-Penrose invariants of the geometry in the Appendix. We adopt conventions such that Greek indices $\alpha, \beta, \gamma \ldots$ run from $0$ to $3$ and Latin indices $i, j, k \ldots$ run from $1$ to $3$ (the three spatial directions). Geometric units $c=8\pi G=1$ are assumed, where c is the speed of light and $G$ is the Newton constant.


\section{The solution}
\label{II}

The properties of the metric we propose here are inspired by the observed homogeneity and isotropy of the universe encoded in the cosmological principle, which is usually put forward by adopting the FLRW model with constant spatial curvature scalar\footnote{A careful analysis of the foundations of the standard cosmological model can be found in \cite{clarkson}, where it is discussed which assumptions can actually be confirmed by observations.}. However, this hypothesis can be relaxed by assuming that just the hypersurfaces at constant time are maximally symmetric without requiring that their corresponding isometries are also symmetries of the whole space-time. With this in mind, the simplest manner to implement this scenario, in the flat case, is through the infinitesimal line element
\begin{equation}
\label{met_f}
ds^2=-e^{f(t,x,y,z)}dt^2+a^2(t)(dx^2+dy^2+dz^2),
\end{equation}
where $f(t,x,y,z)$ is an arbitrary function (at least of class ${\cal C}^2$) and $a(t)$ is the scale factor. With no further assumptions, this is an algebraically general metric of Petrov type I (see Appendix for details). In our approach, we will consider the non-linearity of the Einstein's equations as a whole, thus keeping our analysis in the class of models that cannot be reduced to perturbations of the FLRW spacetimes (see \cite{green,buchert15,green1} and references therein for a fruitful  debate on this subject). Our aim is to describe in a more realistic way how the distribution of matter---through this large scale inhomogeneity---affects the luminosity distance observations trying to deviate as little as possible from the standard model by keeping the homogeneity of the $t={\rm const.}$ hypersurfaces.

The fluid flow will be described by the normalized, irrotational and shear-free vector field $u^\mu=e^{-f/2}\delta^\mu{}_0$ and the flat 3-hypersurfaces, with metric given by $h_{\mu\nu}=g_{\mu\nu}+ u_{\mu}u_{\nu}$, are orthogonal to it. The non-trivial kinematic quantities associated to the four-velocity of the fluid are the expansion coefficient given by
\begin{equation}
\vartheta=u^\mu{}_{;\mu}=3\frac{\dot a}{a}\,e^{-\frac{f}{2}},
\end{equation}
and the acceleration vector $a_\mu=u_{\mu;\nu}u^\nu$ expressed as
\begin{equation}
\label{a_of_f}
a_i=-\frac{1}{2}\,\partial_i f.
\end{equation}
Note that the expansion is not homogeneous and the acceleration is nonzero due to the spatial dependence of $f(t,x,y,z)$. Thus, the energy-momentum tensor can be decomposed with respect to the vector field $u^{\mu}$ into its irreducible parts as
\begin{equation}
T_{\mu\nu}=\rho u_\mu u_\nu+ph_{\mu\nu}+q_{\mu} u_{\nu}+q_{\nu} u_{\mu}+\pi_{\mu\nu},
\end{equation}
where $\rho$ is the energy density, $p$ is the isotropic pressure, $q_\mu$ is the heat flow and $\pi_{\mu\nu}$ is the traceless symmetric anisotropic stress tensor, with $q_\nu\,u^\nu=0$ and $\pi_{\mu\nu}\, u^\nu =0$. The presence of heat flow in this model is well accommodated by the nonzero acceleration of the observers comoving to the fluid in accordance with the Eckart-frame of the thermodynamics of irreversible process \cite{eckart}, where
\begin{equation}
\label{phen_heat}
q_{\mu}=-k a_{\mu},
\end{equation}
with $k$ being the thermal conductivity that could be a function of the thermodynamical variables as temperature, energy density, pressure, among others. On the other hand, the phenomenological and nonrelativistic equation of state for the anisotropic stress ($\pi_{\mu\nu}\propto \sigma_{\mu\nu}$) is not appropriate here, since the comoving observers are shear-free. In this way, if we insist that the viscosity may still be present in this cosmological scenario, then $\pi_{\mu\nu}$ should be related to something else. In the literature, it is possible to find a suitable candidate for it associating the dissipation to the gravitational energy loss, in particular, to tidal forces (see Refs.\ \cite{douce,mimoso,coley94,bss14,bss15} for more details). With this in mind, a possible description could be implemented by considering $\pi_{\mu\nu}$ as a function of $E_{\mu\nu}$, where $E_{\mu\nu}\doteq C_{\mu\alpha\nu\beta} u^{\alpha} u^{\beta}$ is the electric part of the Weyl tensor $C_{\mu\alpha\nu\beta}$, which represents the relativistic extension of tidal forces \cite{ehlers}. Thus, the phenomenological equation of state for the anisotropic stress in the absence of shear would be
\begin{equation}
\label{phen_visc}
\pi_{\mu\nu}=\eta E_{\mu\nu},
\end{equation}
where $\eta$ is a ``tidal'' viscosity coefficient that might also be a function of the thermodynamic variables. After all these considerations, we can compute the EFE in order to determine the geometry of the spacetime.

The $0-0$ component of the EFE, $G_{\mu\nu}=T_{\mu\nu}$, gives
\begin{equation}
\label{coord_trans1}
e^{f(t,x,y,z)}=\fracc{3H^2(t)}{\rho(t,x,y,z)},
\end{equation}
where $H=\dot{a}/a$ and dot means derivative with respect to the time coordinate. For the sake of simplicity, we shall proceed similarly to \cite{leandro} and rewrite the line element using the scale factor as the time coordinate and introduce an explicit dependence in the metric with respect to the energy density through relation (\ref{coord_trans1}), obtaining
\begin{equation}
\label{new_metric}
ds^2=-\fracc{3}{a^2\rho(a,x,y,z)}da^2+a^2(dx^2+dy^2+dz^2).
\end{equation}
Note that the equation $G_{00}=T_{00}$ in this new coordinate system is now an identity. The other components of the EFE lead to
\begin{eqnarray}
\fracc{1}{\sqrt{3\,\rho}} \,\partial_i\rho=q_i&&\label{heat__eisn_eq}\\
\sqrt{\rho}\left[\nabla^2\left(\rho^{-\frac{1}{2}}\right) - \partial_{ij}\left(\rho^{-\frac{1}{2}}\right)\right] - \frac{a^3}{3}\,\frac{\partial\rho}{\partial a} - a^2\rho
=a^2p+\pi_{ij},&& \quad\mbox{for}\ i=j,\label{pres__eisn_eq}\\
\frac{1}{4}\left(\fracc{2\rho\,\partial_{ij}\rho - 3\partial_i\rho \,\partial_j\rho}{\rho^2}\right)=\pi_{ij},&&\quad \mbox{for}\,\, i\neq j,\label{apres__eisn_eq}
\end{eqnarray}
where $\partial_i\equiv\frac{\partial}{\partial x^i}$, $\partial_{ij}\equiv\frac{\partial^2}{\partial x^i\, \partial x^j}$ and $\nabla^2:= \delta^{ij}\partial_{ij}$ is the 3-dimen\-sional Euclidean Laplacian.

If we compare Eq.\ (\ref{heat__eisn_eq}) with Eq.\ (\ref{phen_heat}) using Eqs.\ (\ref{a_of_f}) and (\ref{coord_trans1}), the EFE $G_{0i}=T_{0i}$ set the thermal conductivity to $k=2\sqrt{\rho/3}$. If $\rho(t_0,x,y,z)$ goes to a constant at some limit, then the standard phenomenological equation for the heat flow is recovered \cite{eckart}. Concerning the anisotropic stress, we first compute the electric part of the Weyl tensor in terms of the metric (\ref{new_metric}), obtaining
\begin{equation}
\label{comp_elec}
E_{ij}=\sqrt{\rho}\left[\frac{1}{2}\,\partial_{ij}\left(\rho^{-\frac{1}{2}}\right)-\frac{1}{6}\,\delta_{ij}\,\nabla^2\left(\rho^{-\frac{1}{2}}\right)\right].
\end{equation}
By comparison of Eqs.\ (\ref{phen_visc}) and (\ref{apres__eisn_eq}) with the help of Eq.\ (\ref{comp_elec}), the off-diagonal spatial components of the EFE fix $\eta=-2$. Inserting this information into Eq.\ (\ref{pres__eisn_eq}), the remaining diagonal spatial components of the EFE are all equal to
\begin{equation}\label{Eq:ElipticPDEEnergyDensity}
\frac{\nabla^2(\rho^{-\frac{1}{2}})}{\rho^{-\frac{1}{2}}} =\frac{a^2}{2}\left(a\,\frac{\partial\rho}{\partial a} + 3\gamma\rho\right),
\end{equation}
where, for simplicity, we have assumed a barotropic equation of state of the form $p=(\gamma-1)\rho$ in which $\gamma$ may be considered a function of the scale factor.

The fundamental observers are chosen as those free-falling bodies detecting an almost isotropic CMB radiation, which in our spacetime are discretely distributed along the space instead of being continuously spread out, as in the standard model. In order to clarify this idea, let us consider a time-like geodesic parametrized by the time coordinate $t$. We denote the 3-velocity by $v^i=\tfrac{dx^i}{dt}$ and its Euclidean norm by $v=\sqrt{\delta_{ij}\, v^i\, v^j}$. Using the fact that $\tfrac{dx_\mu}{d\tau}\tfrac{dx^\mu}{d\tau}$ is a constant of motion, we obtain the proper time as
\begin{equation}
\frac{d\tau}{dt} = \sqrt{\frac{e^{f_0}-a_0^2\, v_0^2}{e^{f}-a^2\, v^2}}\, ,
\end{equation}
where $f_0$, $a_0$ and $v_0$ represent the quantities $f$, $a$ and $v$ evaluated along the geodesic at $t=t_0$. Since the coordinate time can be arbitrarily chosen by the transformation $f\to f - \varphi(t)$ and $dt \to e^{\varphi(t)}dt$, we conclude that the cosmological observers we are seeking for are those with $v=0$, for, in this case, they will be free-falling with the fluid and, after choosing $\varphi(t)=f(t,x_0,y_0,z_0)$ along its world-line $(t,x_0,y_0,z_0)$, the cosmic time $t$ will coincide with its proper time $\tau$. In this sense, they are similar to the fundamental observers in the standard model. Moreover, substituting the Friedmann-like equation (\ref{coord_trans1}) into Eq.\ (\ref{a_of_f}), we find that they are exactly located at the spatial critical values of the energy density, since
\begin{equation}\label{Eq:CosmologicalObserversCondition}
\nabla_i f=0 \quad \Longleftrightarrow \quad \nabla_i \rho = 0 \, .
\end{equation}

An overview of the global picture can be put in the following terms: fix a space section of constant time $t_1$ and take those maximum points of $\rho(t_1,\cdot)$ as representing the free-falling (cluster of) galaxies. They should be discretely distributed all over the space in a way that, when averaged on large scales, their number density tends to a constant value $n(t_1)$, and for each world-line representing those cosmological observers, no preferred spatial direction should be allowed. Since it demands a great effort to demonstrate the existence of this global picture, the pursuit of this issue is left to a forthcoming manuscript \cite{bgs2}. At the end of this section we will return to this point, where we give one further evidence in order to justify our heuristic arguments.

One simple manner to ensure the existence of the cosmological observers, that is, the geodesics with $v=0$, is by taking $\rho$ to be a separable function of the time and the space variables, allowing the existence of critical points, such that for those observers the condition (\ref{Eq:CosmologicalObserversCondition}) holds. We can formulate it as
\begin{equation}\label{definicao_Xi}
\rho(t,x,y,z)=\frac{\epsilon(t)}{[\chi(x,y,z)]^2}.
\end{equation}
In this case, the critical values of $\rho(t,\cdot)$ and $\chi$ coincide, implying that they are not time dependent. As a consequence, Eq.\ (\ref{Eq:ElipticPDEEnergyDensity}) can be split in a time dependent part which, in terms of the variable $a$, reads
\begin{equation}
\label{eq_rho_a}
a^3\,\frac{d\epsilon}{da}+3\gamma a^2 \epsilon=-2\kappa ,
\end{equation}
and a spatial counterpart,
\begin{equation}
\label{laplacian}
\chi\nabla^2\chi=- \kappa,
\end{equation}
where $\kappa$ is a constant of separability.

First, let us analyze the dependence of the energy density with respect to the scale factor. Eq.\ (\ref{eq_rho_a}) can be directly integrated in the case of $\gamma$ constant, yielding

\begin{equation}\label{sol_eq_rho_a}
\gamma=const. \quad \Rightarrow \quad
\epsilon(a)=\epsilon_1 \left(\frac{ a_0}{a}\right)^{3\gamma}-\frac{2\kappa}{(3\gamma-2)\, a^2},
\end{equation}
where $\epsilon_1 = \epsilon_0 +\frac{2\kappa}{(3\gamma-2)\,a_0^2}$ and $\epsilon_0 = \epsilon(a=a_0)$. The case $\gamma=2/3$ has to be solved separately, resulting in
\begin{equation}\label{sol_eq_rho_a_gama23}
\gamma=2/3 \quad \Rightarrow \quad
\epsilon(a)=\epsilon_0 \left(\frac{ a_0}{a}\right)^{2}-\frac{2\kappa}{a^2}\,\ln\left(\frac{a}{a_0}\right).
\end{equation}

The first term on the right hand side of Eq.\ (\ref{sol_eq_rho_a}) is the same as in the flat FLRW models and the second term (proportional to $a^{-2}$) mimics a spatial curvature term for $\kappa\neq0$ and $\gamma\neq2/3$ that plays an important role at late times in the evolution of the universe, as we shall see. It should also be noticed that this solution is invariant under the transformation $a\rightarrow \lambda a$ and $\kappa\rightarrow\lambda^2\kappa$. In an expanding universe model there is a combination of the constants such that $\epsilon(a)$ vanishes at a finite value of the scale factor given by
\begin{eqnarray}
\frac{a}{a_0}=\, \left[\frac{2\kappa}{ (3\gamma -2) \,\epsilon_1\,a_0^2}\right]^{\frac{1}{2-3\gamma}},\quad\mbox{for}\quad \gamma\neq\frac{2}{3}\quad\mbox{and}\nonumber\\
\frac{a}{a_0}= \, \exp \left(\frac{\epsilon_0 \, a_0^2}{2 \kappa}\right),\quad\mbox{for}\quad \gamma=\frac{2}{3}.
\end{eqnarray}
When this happens the metric becomes singular in the same manner as the closed FLRW case.

Therefore, by assuming the phenomenological thermodynamic relations and a separation of variables of the form (\ref{definicao_Xi}) with a linear equation of state for the isotropic pressure, the geometry is then completely determined by solutions of Eq.\ (\ref{laplacian}) with appropriate boundary conditions. In the literature, this equation is known as the static version of a nonlinear Klein-Gordon equation whose potential is inversely proportional to $\chi$ \cite{polyanin}. Some solutions of this elliptic partial differential equation (\ref{laplacian}) can be found in implicit form as
\begin{equation}
c_1x + c_2y + c_3z + c_4 +c_5 - \sqrt{c_1^2 + c_2^2 + c_3^2}\int\frac{d\chi}{\sqrt{-2\kappa\ln\chi- (c_1^2 + c_2^2 + c_3^2) c_4}}=0.
\end{equation}
A preliminary analysis indicates that the regularity of these solutions depend on the values taken by the constants, that is, on the boundary conditions. These conditions should, of course, comply with the fact that we observe an almost iso\-tropic background radiation. To that end, however, more important than the spatial variation of the energy density is its time evolution. Since the inhomogeneities contribute with a term that goes with $a^{-2}$, they should become negligible for small values of the scale factor and, therefore, should not spoil the CMB isotropy.

\begin{figure}[htb]
    \centering
    \includegraphics[width=8cm,height=6cm]{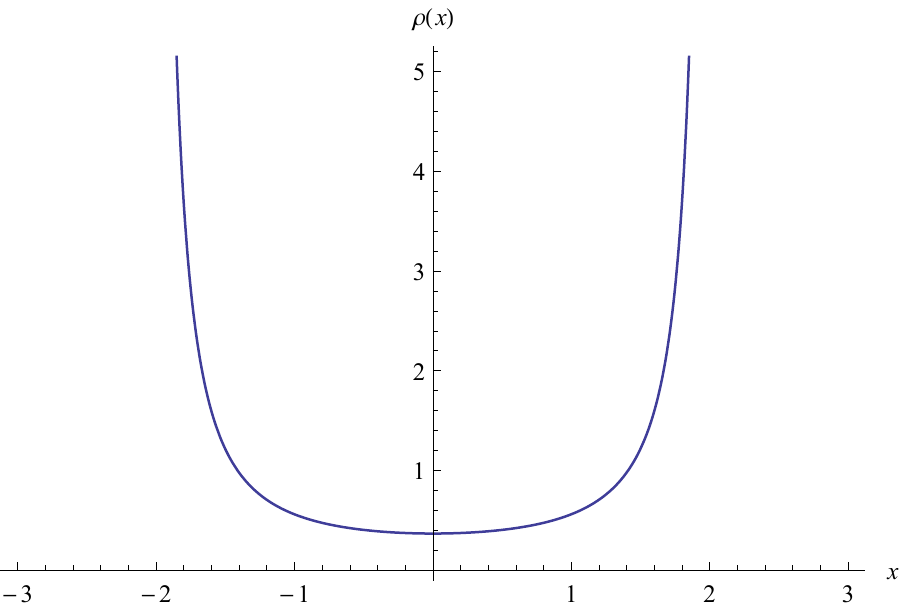}\hspace{1.cm}
 \includegraphics[width=7.5cm,height=6cm]{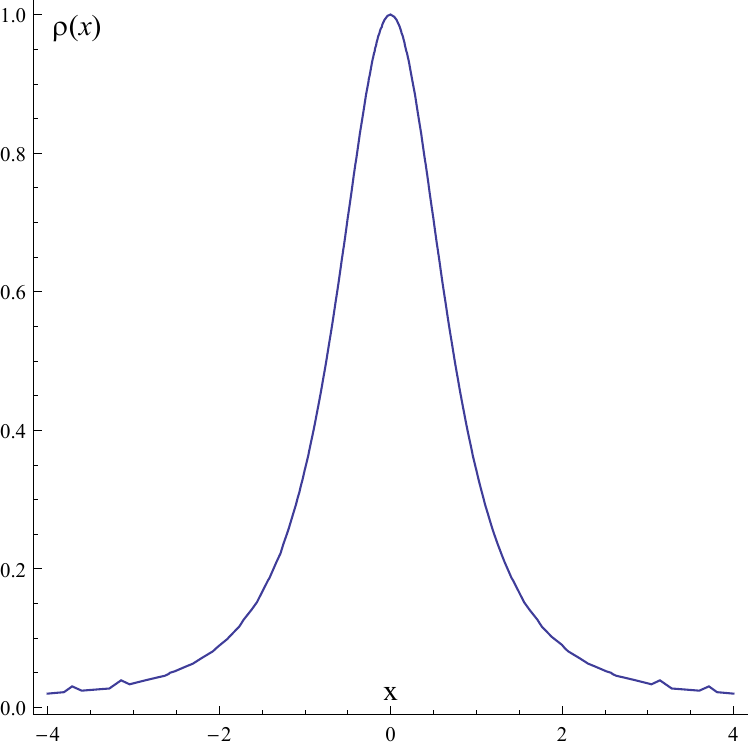}
    \caption{The profiles of $\rho(t_0,x)$ in the 1+1 toy-model, for some $t_0$ fixed. On the left, the typical behavior of the case $\kappa>0$ is shown: the energy density has a minimum at $x=0$ and diverges as $x$ approaches the boundary of the domain of the solution. On the right, the case $\kappa<0$ is illustrated: the peak of the energy density occurs at $x=0$ and then $\rho(t_0,x)\rightarrow0$ when $x\rightarrow\pm\infty$. The cut and glue process of these solutions would eliminate the divergence of the first and the tails of the second, in a way that the global solution is kept smooth.}
    \label{fig:my_label}
\end{figure}

For the sake of illustration, let us consider a toy-model corresponding to the one-dimensional version of Eq.\ (\ref{laplacian}), that is, $d^2\chi(x)/dx^2+\kappa/\chi(x)=0$. This equation admits an analytic solution for $\chi$ in terms of the exponential of the inverse error function and, after some manipulations, the energy density can be written as
\begin{equation}\label{Eq:ExactSolutionToyModel}
\rho(t,x)=\epsilon(t)\frac{e^{2C_1/\kappa}}{(C_2)^2}\times\left\{\begin{array}{ll}
\exp\left\{2
 \left[{\rm Erf}^{-1}(\alpha x + \beta)\right]^2 \right\},&\qquad \kappa>0\\[2ex]
\exp\left\{-2
 \left[{\rm Erfi}^{-1}(\alpha x + \beta)\right]^2 \right\},&\qquad \kappa<0,
 \end{array}\right.
\end{equation}
where $C_1$ and $C_2$ are integration constants. We denote ${\rm Erf}^{-1}(z)$ and ${\rm Erfi}^{-1}(z)$ as the inverse error function and the inverse imaginary error function, respectively. The auxiliary constants $\alpha$ and $\beta$ are
$$
\alpha=\pm\sqrt{\frac{2|\kappa|}{\pi}}\,\frac{e^{-C_1/\kappa}}{C_2},\quad \mbox{and}\quad \beta=\frac{2}{\sqrt{\pi}}\int_0^{\sqrt{\frac{C_1}{|\kappa|}}}\exp\left(-\mbox{sgn}(\kappa)u^2\right)du
$$
The behavior of the above functions are depicted in Fig.\ (\ref{fig:my_label}), where we see that the sign of $\kappa$ plays an important role in the determination of the energy density. For $\kappa>0$, $\rho(t,x)$ is well defined only in the bounded interval $|\alpha x + \beta|\leq1$, a feature inherited from ${\rm Erf}^{-1}(z)$, diverging at the extrema. For $\kappa<0$, $\rho(t,x)$ is defined for all $x\in\mathbb{R}$ going asymptotically to zero when $x\rightarrow\pm\infty$. In principle, both cases are interesting because they possess critical points of the energy density where the fundamental cosmological observers can be defined.  From a heuristic viewpoint, it is straightforward to imagine these two different solutions being cut and glued side by side in order to form a global smooth structure of alternating regions of voids ($\kappa>0$) and concentrations of matter ($\kappa<0$). Even though the reach of the 1+1 model is very limited, it gives a rough idea of a 3+1 model with matter discretely accumulated along the spatial sections. Clearly, the formal mathematical setup requires a more detailed analysis which is left to a forthcoming manuscript \cite{bgs2}, as mentioned before.


\section{Redshift and luminosity distance}
\label{III}

We perform now an analysis of the luminosity distance in this geometry restricted to small values of the redshift. With this in mind, we compute the geodesic equations for the metric (\ref{met_f}), which are given by
\begin{eqnarray}
&&\frac{d k^0}{d\lambda} + H (k^0)^2 + 2\, \frac{\chi'}{\chi}\,k^0 =0,\label{geo1}\\[1ex]
&&\frac{d\vec k}{d\lambda} + \frac{\chi\nabla\chi}{a^2}(k^0)^2 + 2 H\,k^0\,\vec k=0,\label{geo2}
\end{eqnarray}
where $k^0=dt/d\lambda$, $\vec k=\left(\frac{dx}{d\lambda},\frac{dy}{d\lambda},\frac{dz}{d\lambda}\right)$ and $\chi'=\vec k\cdot \nabla\chi$ with $\lambda$ as the affine parameter of the geodesics and where dot ($\cdot$) stands for the usual Euclidean scalar product. The line element provides the constraint
\begin{equation}
\label{first_line_el}
b=-\chi^2(k^0)^{2}+a^2(t) (\vec k\cdot\vec k),
\end{equation}
where $b=-1,0,+1$ if the geodesic is time-like, light-like or space-like, respectively. In particular, to find the corrected expression for the redshift, we only need to study null geodesics. Thus, substitution of Eq.\ (\ref{first_line_el}) into Eq.\ (\ref{geo1}), with $b=0$, yields
\begin{equation}
\label{first_int_geo1}
k^0=\frac{E}{a\chi^2},
\end{equation}
where $E$ is an integration constant.

The redshift is defined as
\begin{equation}
1+z=\frac{(-u_{\mu}k^{\mu})|_{e}}{(-u_{\mu}k^{\mu})|_{o}},
\end{equation}
in which the subscript $e$ indicates the spacetime event where the photon was emitted and the subscript $o$ indicates the spacetime event where the photon was observed. The spatial dependence of $z$ through $\chi$ makes the redshift expression completely different from the one in FLRW models.
In this way, the equation for $z$ becomes

\begin{equation}
\label{z_lambda}
1+z(\lambda) =\frac {a_{0}\chi_0}{a(\lambda)\chi(\lambda)},
\end{equation}
where $a_0$ is the scale factor today and $\chi_0$ is the inverse square root of the spatial energy density at the location in which the photon is observed. In virtue of the freedom in the parametrization of the null geodesics, we can choose $(-u_{\mu}k^{\mu})|_{o}=1$ and this implies $E=a_0\chi_0$ .

Now we proceed to calculate the luminosity distance. First, we consider the Sachs equations \cite{ferreira,bentivegna}
\begin{eqnarray}
&&\frac {d^2D_A}{d\lambda^2} + \left( {\Sigma}^{2}+\frac{1}{2}R_{\mu\nu}k^{\mu}k^{\nu} \right)\,D_A=0,\label{sachs1}\\
&&\frac {d\Sigma}{d\lambda}+2\left( \frac {d\ln D_A}{d\lambda}\right)\Sigma=C_{\alpha\beta\mu\nu}m^{\alpha}k^\beta m^{\mu}k^\nu,\label{sachs2}
\end{eqnarray}
for the angular diameter distance $D_A$ and the shear $\Sigma$ of the null congruence, where $m^\mu$ is a space-like vector orthogonal to $k^\mu$. We set the initial conditions for the observer as
\begin{equation}
D_A(0)=0 \quad{\rm and}\quad\frac{dD_A}{dz}{\bigg|}_0=\frac{1}{H_0} ,
\end{equation}
where $H_0$ is the Hubble parameter at the instant of time in which the photon is observed. Thus, the Sachs equations decouple for $z=0$ and we can proceed with the determination of the luminosity distance using only Eq.\ (\ref{sachs1}). In this case we have $\frac{d^2D_A}{d\lambda^2}=0$, which implies that
\begin{equation}
\label{DerivadaSegundaDA}
\left(\frac{dz}{d\lambda}\right)^{2} \frac{d^2D_A}{dz^{2}} = - \left(\frac{d^{2}z}{d\lambda^{2}}\right)\frac{1}{H_0}   \qquad \mbox{for $z=0$}.
\end{equation}
In terms of the affine parameter $\lambda$, this equation is, of course, the same as the one obtained for FLRW models. However, the presence of inhomogeneities leads to a completely different equation in terms of redshift due to relation (\ref{z_lambda}), whose consequences we shall see soon. Even for small $z\neq0$, the analysis is more intricate and even though the hypothesis of neglecting the Weyl lensing could be helpful, the determination of the angular diameter distance is still not straightforward due to the inhomogeneities in the Ricci term of Eq.\ (\ref{sachs1}).

Following the steps presented in Ref.\ \cite{villani}, we consider Taylor expansions of $a$ and $\chi$ up to second order in $\lambda$ as
\begin{equation}
a(\lambda) = a_0\left(1 + a_1H_0\lambda + \frac{1}{2}\, a_2H_0^2\,\lambda^2\right) + O(\lambda^3)
\end{equation}
and
\begin{equation}
\label{taylor_chi}
\chi(\lambda)=\chi_0\left(1+\chi_1H_0\lambda+\frac{1}{2}\, \chi_2H_0^2\,\lambda^2\right) + O(\lambda^3).
\end{equation}
Using Eqs.\ (\ref{geo1}) and (\ref{first_int_geo1}), we see that the parameters $a_1$ and $a_2$ are related to the constants $\chi_0$ and $\chi_1$ through
\begin{equation}
a_1=\frac{1}{\chi_0}\qquad\mbox{and}\qquad a_2=-\frac{1+q_0+2\chi_0\chi_1}{\chi_0^2},
\end{equation}
where $q(t)=-\ddot a/aH^2$ is the deceleration parameter as defined for FLRW metrics and $q_0$ stands for its value at the moment of the observation. At this order, we are able to compute the contribution of the inhomogeneity to the deceleration parameter and then try, in a first moment, to put bounds on such contributions based on observations and compare our predictions with some cosmological models commonly discussed in the literature.

Thus, differentiating Eq.\ (\ref{z_lambda}) with respect to $\lambda$ and rewriting the outcome in terms of $z$ yields

\begin{equation}\label{DerivadasDAZ0}
\frac{dz}{d\lambda}\Big|_{z=0} = -H_0\, (a_1+\chi_1)
\quad \mbox{and}\quad
\frac{d^2z}{d\lambda^2}\Big|_{z=0}=H_0^2\,[2(a_1^2+a_1\chi_1 + \chi_1^2)-\chi_2-a_2].
\end{equation}
As we use the relation $D_L=(1+z)^2D_A$ between the luminosity and angular diameter distances \cite{Ether}, the  Taylor expansion of order $z^2$ is expressed as \cite{visser}
\begin{equation}
D_X=\frac{D_X^{(1)}}{H_0}\,z+\frac{D_X^{(2)}}{2H_0}\,z^2 , \qquad \mbox{with}\quad X=A,L\ ,
\end{equation}
where the coefficients are linked through
\begin{equation}
\begin{array}{lcl}
D_L^{(1)}&=&D_A^{(1)}=1,\\[1ex]
D_L^{(2)}&=&D_A^{(2)}+4\equiv 1-q_{{\rm eff}}
\end{array}
\end{equation}
with $q_{{\rm eff}}$ as the effective (observed) deceleration parameter. We get from Eqs.\ (\ref{DerivadaSegundaDA}) and (\ref{DerivadasDAZ0}) that
\begin{equation}
 D_A^{(2)}=-\frac{2(a_1^2+a_1\chi_1 + \chi_1^2)-\chi_2-a_2}{(a_1+\chi_1)^2},
\end{equation}
leading to the following expression for the effective deceleration parameter
\begin{equation}
\label{q_eff}
q_{{\rm eff}}=\frac{q_0 + 2\chi_1 - 4\chi_0\chi_1 - \chi_0^2\chi_1^2 -\chi_0^2\chi_2}{(1+\chi_0\chi_1)^2}.
\end{equation}
It should be noticed that there are two different contributions from inhomogeneities to $q_{{\rm eff}}$: first, we have the local inhomogeneities surrounding the observer that affect the measure of the deceleration parameter whenever $\chi_1,\chi_2\neq0$; the other contribution from the inhomogeneities is global and comes from the ``effective curvature'' term proportional to $\kappa$ that changes the value of $q_0$ independently of position in space. Note also that $q_{{\rm eff}}$ reduces to the usual FLRW $q_0$ value when we neglect the inhomogeneous parameters $\chi_1$ and $\chi_2$ (and higher orders), since the only possible value for $\kappa$ in this limit is zero.


\section{Contributions to the deceleration parameter from inhomogeneities}
\label{IV}

We now provide a preliminary analysis showing how our model can accommodate the predictions regarding both the standard $\Lambda$CDM and LTB inhomogeneous models. To that end, we consider the physical interpretation of the parameters composing $q_{{\rm eff}}$, where we let $(t_0,x_0,y_0,z_0)$ represent the space-time event of observation at $z=0$ and set $f(t_0,x_0,y_0,z_0)=0$, $\rho_0:=\rho(t_0,x_0,y_0,z_0)$ and  $\epsilon_0=\epsilon(a_0)=\rho_0$. This implies that
\begin{equation}\label{Xi0}
\chi_0 = 1 \quad \mbox{,} \quad E=a_0 \quad \mbox{and} \quad  \epsilon_0=\rho_0=3H_0^2 \, .
\end{equation}
With $\epsilon_0$ and $\gamma_0=\gamma(a_0)$  in Eq.\ (\ref{eq_rho_a}), we determine $q_0$ as
\begin{equation}
q_0 = \frac{3}{2}\, \gamma_0 -1 +  \frac{\kappa}{3\,a_0^2\, H_0^2} \, .
\end{equation}
Note that even $q_0$ has a contribution due to inhomogeneities through $\kappa$. From Eqs.\ (\ref{first_line_el}) and (\ref{first_int_geo1}) we have for the norm of the 3-dimensional wave vector $\|\vec k\| = \sqrt{\vec k\cdot\vec k}$
\begin{equation}
\|\vec k\| =  \frac{1}{a_0} \left(\frac{a_0}{a}\right)^2\, \sqrt{\frac{\rho}{\epsilon}} \, .
\end{equation}
As we use the Taylor expansion (\ref{taylor_chi}) for $\chi$ and the definition of $\chi_1$, we obtain
\begin{equation}\label{Formula_X1}
\chi_1 \equiv \left( \frac{1}{H \, \chi}  \, \frac {d\chi}{d\lambda}\right)_{z=0} = -\left(\frac{1}{2 H \,\rho} \, \vec k \cdot \vec{\nabla}\rho \right)_{z=0}= -\frac{1}{6\, a_0 \, H_0^3} \, \|\vec{\nabla}\rho\|_0 \, \cos \psi_0 .
\end{equation}
where we denote by $\psi_0$ the angle between $\vec{\nabla}\rho$ and $\vec k$ at the event of observation and $\|\vec{\nabla}\rho\|_0=\|\vec{\nabla}\rho(t_0,x_0,y_0,z_0)\|$. The definition of $\chi_2$, in its turn, is
\begin{equation}\label{definicao_X2}
\chi_2 \equiv \left( \frac{1}{H^2 \, \chi}  \, \frac {d^2\chi}{d\lambda^2}\right)_{z=0} =
\frac{1}{H_0^2} \left(
\frac{d\vec k}{d \lambda}\cdot \vec\nabla \chi + k^i k^j \,\partial_i\partial_j \chi
\right)_{z=0} .
\end{equation}
If we use Eqs.\ (\ref{geo2}), (\ref{first_int_geo1}) and (\ref{Formula_X1}), we obtain for the first term on the right-hand side
\begin{equation}\label{CalculoX2_1}
\frac{1}{H_0^2} \left(
\frac{d\vec k}{d \lambda}\cdot \vec\nabla \chi
\right)_{z=0} = \frac{\|\vec{\nabla}\rho\|_0}{3\, a_0 \, H_0^3} \, \cos \psi_0 - \left(\frac{\|\vec{\nabla}\rho\|_0}{6\, a_0 \, H_0^3} \, \right)^2 .
\end{equation}
Concerning the term $\partial_i\partial_j \chi$, we can decompose it into its trace $\Delta \chi$ and its traceless part $B_{ij}$ both at the event of observation. Therefore, from Eq.\ (\ref{laplacian}), we get
\begin{equation}\label{CalculoX2_2}
\partial_i\partial_j \chi (x_0,y_0,z_0) = - \frac{\kappa}{3} \delta_{ij} + B_{ij} .
\end{equation}
Since the last term becomes diagonal by an Euclidean rotation, we can take it as the diagonal matrix
$$(B_{ij})= B \, \textbf{diag} \{\sin b, \sin (b + 2 \pi/3), \sin (b + 4 \pi/3) \},$$
where $B$ measures the intensity of the anisotropy in the Hessian matrix of $\chi$ at the observation point and $b$ its angular distribution.
Defining the polar and azimuthal angles in the usual way, with $k^1=\|k \| \cos \varphi$, $k^2=\|k \| \cos \theta \sin \varphi$ and $k^3=\|k \| \sin \theta \sin \varphi$, we substitute Eqs.\ (\ref{CalculoX2_1}) and (\ref{CalculoX2_2}) in (\ref{definicao_X2}) to get
\begin{eqnarray}
\chi_2 = -  \, \frac{\kappa}{3\,a_0^2\, H_0^2}  + \frac{\|\vec{\nabla}\rho\|_0}{3\, a_0 \, H_0^3}\, \cos \psi_0 -  \left(\frac{\|\vec{\nabla}\rho\|_0}{6\, a_0 \, H_0^3} \, \right)^2+ \frac{B}{a_0^2\, H_0^2} \, A(\varphi, \theta,b)  ,
\end{eqnarray}
where
\begin{equation}
A(\varphi, \theta,b)= \left( 1 - \frac{3}{2}\sin^2 \varphi \right)\, \sin b - \frac{\sqrt{3}}{2}\, \cos(2\theta)\, \sin^2 \varphi\,  \cos b  .
\end{equation}
Collecting all these terms and putting them back into the effective deceleration parameter given by Eq.\ (\ref{q_eff}), yields
\begin{equation}
\label{q_eff_qs}
q_{{\rm eff}}=\frac{1}{\left(1-\frac{\|\vec{\nabla}\rho\|_0\,\cos\psi_0}{6\, a_0 \, H_0^3}\right)^2}\left[\frac{3}{2}\gamma_0 -1 +  \frac{2\kappa}{3\,a_0^2\, H_0^2} + \left(\frac{\|\vec{\nabla}\rho\|_0\,\sin \psi_0}{6\, a_0 \, H_0^3} \, \right)^2 \, - \frac{B}{a_0^2\, H_0^2} \, A(\varphi,\theta,b)\right].
\end{equation}
Therefore, the general expression for the effective deceleration parameter in our model is inhomogeneous and anisotropic, since it is sensitive to the spatial gradients of the energy density and the directions from the incoming photon through the angles $\theta$ and $\varphi$.

Let us make an extra hypothesis and assume that the observation point lies in what could be considered a critical point of the energy density, a local maximum or minimum depending on whether it represents a matter clump or a void, respectively, such that $\|\vec{\nabla}\rho\|_0$ vanishes. Besides, we can assume that observations are typically performed in many directions so that it is reasonable to consider a redshift-distance relation as the average made over all directions on the sky, in a first approximation.
Thus, we can eliminate the angular dependence from Eq.\ (\ref{q_eff_qs}) so that the effective deceleration parameter reduces to
\begin{equation}
q_{{\rm eff}}=\frac{3}{2}\gamma_0 -1 +  \frac{2\kappa}{3\,a_0^2\, H_0^2}\, .
\end{equation}

In order to compare our inhomogeneous model with some other cosmological scenarios, we first set the dimensionless inhomogeneous constant $\Omega_I$ to be
\begin{equation}
 \Omega_I = -\, \frac{2 \, \kappa}{3\, H_0^2 \, a_0^2} \, .
\end{equation}
It becomes clear from Eq.\ (\ref{eq_rho_a}) that the ratio $\kappa/a_0^2$ is invariant under scale changes in $a(t)$, and therefore it is invariant just like $H_0$. Under the suitable assumption $\gamma=\gamma(a)$ in Eq.\ (\ref{eq_rho_a}), we can write the homogeneous part of the energy density as a sum of contributions of matter ($\Omega_M$) and an effective contribution from the inhomogeneity $(\Omega_I)$, obeying the constraint $\Omega_M + \Omega_I = 1$. The effective deceleration parameter can then be rewritten as
\begin{equation}\label{Eq:qeff_Omegas}
 q_{\rm eff} = \frac{1}{2} \left(\, \Omega_M - \Omega_I \,\right)= \Omega_M -\frac{1}{2}\, .
\end{equation}
From this we see that $-0.5$ is the lower bound for $q_{\rm eff}$ in order that the matter content $\Omega_M$ be positive. If we compare this equation with the standard flat FLRW models\footnote{We consider a reconstructed value as it does not assume any specific matter content {\it a priori} and that is obtained using a kinematic approach in a model independent way. See \cite{sandro} for more details.}, assuming $q_{\rm eff}=-0.5\pm 0.1$, then there is a possible interval $-0.5<q_{\rm eff}\leq-0.4$ for which $\Omega_M\geq0$. Then, from Eq.\ (\ref{Eq:qeff_Omegas}), we see that the maximum value of the density parameter of matter would correspond to 0.1 of the energy content of such universe and, consequently, we should be in a local maximum of the energy density ($\nabla^2\rho(t_0, x_0, y_0, z_0)<0$) since $\Omega_I>0$ according to the constraint. Thus, there is a narrow window where we can explain these observations within our model without extra components for the cosmic fluid. On the other hand, if we compare our expression for $q_{\rm eff}$ with some LTB inhomogeneous models, our proposal becomes more realistic. Recently, a special class of LTB models where the effective deceleration parameter is positive at low redshifts \cite{pascual,alnes,vander,garcia,feb,rossi,tovar} has been confronted with observations. Using their predictions in our model, it is possible to have scenarios where $\Omega_M$ increases above $0.5$, namely, $q_{\rm eff}>0$ from Eq.\ (\ref{Eq:qeff_Omegas}) and then the contribution of $\Omega_I$ to the late evolution of the Universe is reduced without severe restrictions on $\kappa$. Thus, our solution could accommodate small inhomogeneities that however can affect the late time evolution of the universe. In this vein, this model might provide a reasonable explanation to the apparent acceleration of the Universe respecting homogeneity and isotropy as the FLRW models, but in the intrinsic symmetric sense.


\section{Concluding Remarks}\label{V}
An exact solution of the EFE that presents maximally symmetric submanifolds corresponding to tri-dimensional flat spaces of constant time was derived. The matter content required for this is a viscous fluid where the dissipative terms, i.e., the heat flow and the anisotropic stress are given through physically reasonable equations of state involving the acceleration of the comoving observers and the electric part of the Weyl tensor, respectively. The set of equations that come out from the EFE can be split into time and space, where the time evolution is driven by a Friedmann-type equation with an effective curvature term ($\sim a^{-2}$) while the spatial dependence is given by a of nonlinear Klein-Gordon equation. In comparison to the FLRW models, the expressions for the redshift and the luminosity distance are more complicated due to the presence of inhomogeneities. Thus, for very small redshifts, we solve the Sachs equation perturbatively up to second order admitting a Taylor expansion of the solution. In this way, we were able to compute the angular diameter and luminosity distances and, consequently, find an equation for the cosmological deceleration parameter which is inhomogeneous and anisotropic, in general. For the sake of simplicity, the location of photon detection was chosen to be a critical point of the energy density and we consider a luminosity distance-redshift relation averaged over all directions such that we can neglect all terms due to local inhomogeneities and anisotropies at this point. Thus, we are left only with the global contribution of inhomogeneities encoded in $\kappa$ to $q_{\rm eff}$. After comparison with both $\Lambda$CDM and LTB scenarios, it turns out that our model favors the fully inhomogeneous cosmological model where $\Omega_M$ can be the dominant content of the universe.

In order to have a complete physically acceptable cosmology, we should be able to obtain an almost iso\-tropic setting for large enough redshifts to explain the cosmic microwave background radiation. Although, at this stage, the behaviour $a^{-2}$ in the time evolution guarantees that the inhomogeneities will be washed out for small values of the scale factor, a careful analysis would demand a suitable choice of the boundary conditions for the full solution and it would be interesting to further investigate and search for predictions considering a perturbative analysis of our metric, which is subject of future work.

Here we have shown, above all other possible conclusions, that different models with nonlinear effects of inhomogeneity (and possibly anisotropy) beyond small perturbations of FLRW must be pushed forward and tested against the observational data. Just after such a tenacious scrutiny we could conclude that the Universe is of a $\Lambda$CDM type, or composed just by ordinary matter in an inhomogeneous way or somewhere in between.

\appendix

\section{Newman-Penrose (NP) invariants of the geometry}
In this section, we calculate all the NP invariants in order to show that the metric (\ref{new_metric}) is indeed an algebracially general space-time. Hence, using the $(a,x,y,z)$ coordinate system, we define a null tetrad basis $(l_+^\mu,l_-^\nu,m^\mu,\bar m^{\mu})$ given by
\begin{equation}
l_{\pm}^{\mu}=\frac{1}{\sqrt{2}}\left(a\sqrt{\frac{\rho}{3}},0,\pm\frac{1}{a},0\right),\quad m^{\mu}=\frac{1}{\sqrt{2}a}\left(0,1,0,i\right),
\end{equation}
with $\bar m^{\mu}$ being the complex conjugate of $m^\mu$. This basis satisfies the relations $l_{\pm}^{\mu}m_{\mu}=0$, $l_{\pm}^{\mu}l^{\mp}_{\mu}=-1$ and $m^{\mu}\bar m_{\mu}=1$. In possession of this, a direct calculation provides the NP scalars associated to the Weyl tensor as

\begin{equation}
\begin{array}{l}
\Psi_0=\fracc{\left(\partial_z + i\partial_x\right)^2\chi}{2a^2\chi},\quad \Psi_1=-\fracc{\sqrt{2}\,\partial_y(\partial_z + i \partial_x)\chi}{4a^2\chi},\quad \Psi_2=\fracc{\nabla^2\chi}{6a^2\chi}-\frac{(\partial_z +i\partial_x)(\partial_z -i\partial_x)\chi}{4a^2\chi},\\[2ex]
\Psi_3=\fracc{\sqrt{2}\,\partial_y(\partial_z - i \partial_x)\chi}{8a^2\chi},\quad
\Psi_4=\fracc{\left(\partial_z - i\partial_x\right)^2\chi}{8a^2\chi}.
\end{array}
\end{equation}
For completeness, the invariants related to the Ricci tensor are
\begin{equation}
\begin{array}{l}
\Phi_{00}=\fracc{(\partial_z +i\partial_x)(\partial_z -i\partial_x)\chi}{2a^2\chi}-\fracc{1}{\chi^2}\left(\fracc{\ddot a}{a}-\fracc{2\dot a\partial_y\chi}{a^2}-\fracc{\dot a^2}{a^2}\right),\quad \Phi_{01}=\fracc{\sqrt{2}\,(2\dot a- \chi\partial_y)(\partial_z + i \partial_x)\chi}{4a^2\chi^2},\\[2ex]
\Phi_{02}=-\fracc{\left(\partial_z + i\partial_x\right)^2\chi}{4a^2\chi},\quad
\Phi_{11}=\fracc{\partial_{yy}\chi}{4a^2\chi}+\fracc{1}{4\chi^2}\left(\fracc{\dot a^2}{a^2} - \fracc{\ddot a}{a}\right),\quad \Phi_{12}=\fracc{\sqrt{2}\,(2\dot a + \chi\partial_y)(\partial_z + i \partial_x)\chi}{8a^2\chi^2},\\[2ex]
\Phi_{22}=\fracc{(\partial_z +i\partial_x)(\partial_z -i\partial_x)\chi}{8a^2\chi}-\fracc{1}{4\chi^2}\left(\fracc{\ddot a}{a}+\fracc{2\dot a\partial_y\chi}{a^2}-\fracc{\dot a^2}{a^2}\right),\quad R=\fracc{6}{\chi^2}\left(\fracc{\ddot a}{a} +\fracc{\dot a^2}{a^2}\right) -\fracc{2\nabla^2\chi}{a^2\chi}.
\end{array}
\end{equation}
Following the algorithm presented in \cite{acevedo} it is straightforward to check that the metric (\ref{new_metric}) is a Petrov Type-I spacetime.

\acknowledgments
The authors are in debts with the anonymous Referee for the valuable comments on a previous version of this manuscript. G. B. Santos is supported by CAPES (Brazil) through the grant 88882.317979/2019-1.

\end{document}